\documentclass[journal, hidelinks]{IEEEtran}
\usepackage{amsfonts}
\usepackage{algorithmic}
\usepackage{algorithm}
\usepackage{array}
\usepackage[caption=false,font=normalsize,labelfont=sf,textfont=sf]{subfig}
\usepackage{textcomp}
\usepackage{stfloats}
\usepackage{url}
\usepackage{verbatim}
\usepackage{graphicx}
\usepackage{lipsum}
\usepackage{cite}
\usepackage{orcidlink}
\usepackage{amsmath,amssymb,mathtools}
\usepackage[english]{babel}
\usepackage{caption}
\usepackage{bm}
\usepackage{csquotes}
\usepackage[acronym,shortcuts]{glossaries}
\usepackage{xcolor}
\usepackage{multicol}

\newacronym{MIMO}{MIMO}{multiple input multiple output}
\newacronym{BS}{BS}{base station}
\newacronym{ML}{ML}{machine learning}
\newacronym{UE}{UE}{user equipment}
\newacronym{RIS}{RIS}{reconfigurable intelligent surface}
\newacronym{UPA}{UPA}{uniform planar array}
\newacronym{ISAC}{ISAC}{intelligent surface-assisted communications}
\newacronym{AoA}{AoA}{angle of arrival}
\newacronym{SINR}{SINR}{signal-to-interference-plus-noise ratio}
\newacronym{AP}{AP}{access point}
\newacronym{CPU}{CPU}{central processing unit}
\newacronym{CF}{CF}{cell-free}
\newacronym{mMIMO}{mMIMO}{massive MIMO}

\newacronym{QC}{QC}{quantum computing}
\newacronym{B6G}{B6G}{beyond-sixth-generation}
\newacronym{HOBO}{HOBO}{higher-order binary optimization}
\newacronym{HUBO}{HUBO}{higher-order unconstrainted binary optimization}
\newacronym{QA}{QA}{quantum annealing}
\newacronym{QML}{QML}{quantum machine learning}
\newacronym{QAOA}{QAOA}{quantum approximated optimization algorithms}
\newacronym{VQA}{VQA}{variational quantum algorithms}
\newacronym{QMO}{QMO}{quantum manifold optimization}

\begin{document}

\title{Quantum Manifold Optimization: A Design Framework for Future Communications Systems}

\author{Getuar Rexhepi\textsuperscript{\orcidlink{0009-0002-3268-522X}},~\IEEEmembership{Student~Member,~IEEE}, \,Hyeon Seok Rou\textsuperscript{\orcidlink{0000-0003-3483-7629}},~\IEEEmembership{Member,~IEEE}, \\Giuseppe Thadeu Freitas de Abreu\textsuperscript{\orcidlink{0000-0002-5018-8174
}},~\IEEEmembership{Senior~Member,~IEEE}
\thanks{G.~Rexhepi, H.~S.~Rou, and G.~T.~F. de~Abreu are with the School of Computer Science and Engineering, Constructor University Bremen, Campus Ring 1, 28759 Bremen, Germany (email: [grexhepi, hrou, gabreu]@constructor.university).}  
\vspace{-4ex}
}
\maketitle

\begin{abstract}
Inspired by recent developments in various areas of science relevant to quantum computing, we introduce \emph{quantum manifold optimization} (\acs{QMO}) as a promising framework for solving constrained optimization problems in next-generation wireless communication systems.
We begin by showing how classical wireless design problems — such as pilot design in \ac{CF}-\ac{mMIMO}, beamformer optimization in gigantic \ac{MIMO}, and \ac{RIS} phase tuning — naturally reside on structured manifolds like the Stiefel, Grassmannian, and oblique manifolds, with the latter novelly formulated in this work. 
Then, we demonstrate how these problems can be reformulated as trace-based quantum expectation values over variationally-encoded quantum states.
While theoretical in scope, the work lays a foundation for a new class of quantum optimization algorithms with broad application to the design of future \ac{B6G} systems.
\end{abstract}


\glsresetall

\vspace{-2ex}
\section{Introduction}

\Ac{QC} has recently emerged as a transformative technology capable of solving problems that surpass many capabilities of classical computing. 
Leveraging fundamental quantum phenomena such as superposition and entanglement, quantum computing offers revolutionary computational strategies, particularly critical in addressing optimization challenges within the demanding landscape of \ac{B6G} wireless communications. 
These future wireless networks necessitate ultra-high data rates, massive device connectivity, and stringent latency requirements, posing optimization problems of unprecedented complexity \cite{Gyongyosi2022quantum,Zhao2024Quantum,Yukiyoshi_2024}.

\begin{figure*}
\vspace{-2ex}
\centering
\includegraphics[width=0.90\textwidth]{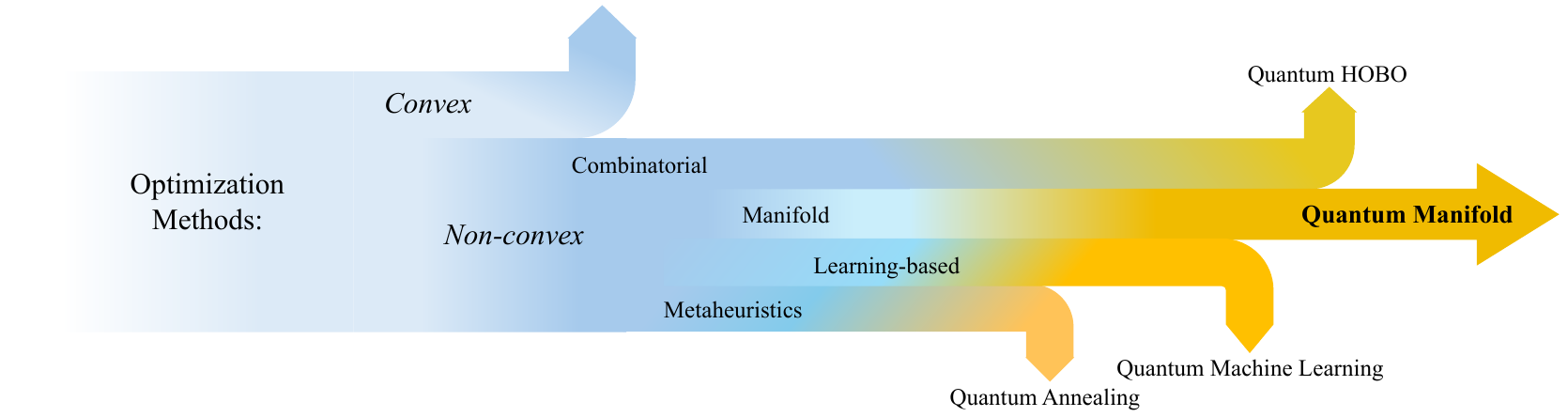}
\vspace{1ex}
\caption{Timeline of the evolution of optimization theory and algorithms towards quantum manifold methods.}
\end{figure*}

To address these challenges of the \ac{B6G}, several quantum optimization methodologies have gained prominence \cite{Abbas2024challenges}, such as \ac{HOBO} approaches which reformulate and encode complex optimization scenarios into higher-order binary decision variables. 
However, such formulations often encounter scalability issues due to intricate constraints and the complexity of encoding these higher-order interactions into quantum systems \cite{Norimoto2023qubo, Jun2023hubo}. 

Alternatively, \ac{QA}, enabled by commercial quantum platforms like D-Wave, has shown also promise for solving NP-hard combinatorial problems by performing heuristic optimization by exploring energy landscapes to minimize objective functions. 
Despite its practical success in small-to-medium scale scenarios, \ac{QA} also suffers from inherent limitations, including discretization of continuous parameters, hardware-induced noise, limited coherence times, and embedding complexities when mapping real-world problems onto quantum hardware \cite{Kim2021heuristic}.

On the other hand, there are methods also classified under \ac{QML}, which encompass various techniques such as \ac{QAOA}. 
These approaches aim to enhance classical \ac{ML} techniques by integrating quantum computation, potentially offering quantum-enhanced learning capabilities and parallel processing benefits. 
Nevertheless, \ac{QML} methods grapple with significant hurdles such as quantum state preparation challenges, barren plateau phenomena in optimization landscapes, and limited circuit depth imposed by current quantum hardware constraints \cite{Narottama2023qml}.

In response to the limitations of current quantum optimization methods, \ac{QMO} has recently been investigated in the physics and chemistry communities, leveraging the intrinsic geometric structures of optimization problems to directly perform optimization on manifolds, such as Stiefel and Grassmannian manifolds, with natural constraints like orthogonality or constant modulus \cite{Smart2024manybody}.
The \ac{QMO} framework integrates the rigorous mathematical properties of manifold optimization with the intrinsic advantages of quantum computation, in addition to directly accommodating geometric constraints without binary discretization, significantly reducing computational overhead, unlike conventional quantum optimization and algorithmic methods.

In turn, it is well-known that general manifold optimization \cite{boumal2023,Absil2008manifold} demonstrate considerable efficacy in classical wireless communication applications, including the design of beamforming precoding matrices, optimization of phase shifts in \ac{RIS}, and subspace estimation, providing superior solution quality and computational efficiency over traditional unconstrained methods \cite{Absil2008manifold,Chepuri2023integrated}.

In light of the above, this article aims to introduce \acf{QMO} as a pioneering framework specifically tailored for future \ac{B6G} wireless communications. 
By establishing a theoretical foundation and general formulation which can be easily projected to many applications in wireless communications, we propose \ac{QMO} as an essential methodology poised to revolutionize wireless system design.

In addition to describing how the contributions of \cite{Smart2024manybody} can be built upon towards solving problems of interest in communications systems design involving the Stiefold and Grassmanian manifolds, we also contribute a novel \ac{QMO}-formulation enabling optimization over Oblique Manifolds, exemplified by the pilot allocation problem, which serves as an example that the our discussion has the potential to go far beyond what is offered in this first article.

\newpage

\section{From Classic Manifolds \\ to Quantum Manifold Optimization}
Manifold optimization techniques emerged as a powerful tool for solving optimization problems, whose solutions inherently lie on a manifold -- a space that locally resembles Euclidean space but has a more complex global structure.
In addition to the success in many other areas of science and various applications, manifold optimization has been successfully applied in wireless communications, particularly in context of large-scale \ac{MIMO} systems, \acf{RIS} optimization, and waveform design for \ac{ISAC}, where the main advantage is its ability to handle the inherent constraints and non-convexities.

Building on top of the literature of manifold optimization techniques and theory, the authors in \cite{Smart2024manybody} have recently proposed a novel framework that seamlessly integrates manifold optimization techniques with quantum state representations and consequently quantum optimization.
Specifically, the many-body eigenstate problem is reformulated as an optimization problem over Riemannian manifolds, with the proposed method focusing on \emph{Stiefel} and \emph{Grassmannian} manifolds, which intrinsically enforce the orthogonality constraints necessary for capturing multiple eigenstates, thereby providing an efficient pathway for quantum computations.

The Stiefel manifold, $\mathrm{St}(n,p)$, is defined as:
\begin{equation}
\mathrm{St}(n,p) = \{\mathbf{X} \in \mathbb{R}^{n \times p} : \mathbf{X}^{\mathrm{T}} \mathbf{X} = \mathbf{I}_p\},
\end{equation}
where $ \mathbf{I}_p $ denotes the $ p \times p $ identity matrix. 
Optimization problems on the Stiefel manifold frequently occur in applications requiring orthogonality constraints, such as beamforming matrix design and precoding in massive MIMO systems.

The Grassmannian manifold, $ \mathrm{Gr}(n,p) $, captures equivalence classes of $ p $-dimensional subspaces of $ \mathbb{R}^n $, defined via:
\begin{equation}
\mathrm{Gr}(n,p) = \mathrm{St}(n,p)/ \sim,
\end{equation}
where the equivalence relation \( \sim \) identifies matrices whose columns span the same subspace. 

This manifold naturally arises in problems such as subspace estimation and interference alignment, and it has recently been shown \cite{Smart2024manybody} that optimization on such manifolds can be elegantly represented via trace-based formulations.

Specifically, for the Stiefel manifold optimization, the many-body eigenstate problem is formulated as
\begin{equation}
\min_{\mathbf{X} \in \mathrm{St}(n,p)} \quad f_{\mathrm{St}}(\mathbf{X}) = \tfrac{1}{2}\mathrm{Tr}(\mathbf{X}^{\mathrm{T}} \mathbf{H} \mathbf{X} \mathbf{K}),
\end{equation}
where $\mathbf{H}$ is a Hermitian matrix representing the system Hamiltonian, and $K$ is typically diagonal, encoding eigenvalue ordering, which for the Grassmannian manifold reduces to
\begin{equation}
\min_{\mathbf{X} \in \mathrm{Gr}(n,p)} \quad f_{\mathrm{Gr}}(\mathbf{X}) =  \tfrac{1}{2}\mathrm{Tr}(\mathbf{X}^{\mathrm{T}} \mathbf{H} \mathbf{X})
\end{equation}
reflecting a simpler optimization landscape while maintaining critical geometric constraints.

Then, \cite{Smart2024manybody} elaborates on the representation of the objectives under the \ac{QMO} framework, centered on calculating Riemannian gradients, performing retractions, and transporting tangent vectors.
However, these techniques are not limited solely to the Stiefel and Grassmannian manifolds, can be extended to any Riemannian manifold where the inherent geometry is beneficial for solving constrained optimization problems.

Therefore, in the following, we novelly derive the \ac{QMO} formulations for the oblique manifold,  appearing in critical wireless problems such as pilot design, as an exemplary case to illustrate the reformulation procedure of classic manifold optimization problem into the \ac{QMO} framework.
In addition, in the next section, a series of relevant use cases in wireless communications system designs are considered, and the \ac{QMO}-based problem formulations are derived.

\vspace{-1ex}
\subsection{The Oblique Manifold}

Consider the \emph{oblique manifold}, defined as
\begin{equation}\label{eq:obliqu_man}
\mathcal{OB}(n, d) = \left\{ \mathbf{X} \in \mathbb{C}^{n \times d} \mid \operatorname{diag}(\mathbf{X}^\dagger \mathbf{X}) = \mathbf{I}_d \right\},
\end{equation}
which represents the Cartesian product of $d$ unit spheres in $\mathbb{C}^n$, where each column of $\mathbf{X}$ lies on the complex unit sphere. 

Unlike the Stiefel manifold, where columns must be orthogonal, the oblique manifold permits independent unitary transformations per column. 
This flexibility suits applications in wireless communications, especially when $n < d$, with $d$ representing the space dimension and $n$ the number of columns.
The tangent space at a point $\mathbf{X}$ on the oblique manifold comprises of all infinitesimal directions that maintain manifold membership, namely
\begin{equation}\label{eq:oblique_tan}
\mathcal{T}_\mathbf{X}(\mathcal{OB}) = \left\{ \mathbf{Z} \in \mathbb{C}^{n \times d} \,\middle|\, \operatorname{diag}(\mathbf{X}^\dagger \mathbf{Z}) = \mathbf{0} \right\},
\end{equation}
where $\operatorname{diag}(\cdot)$ extracts the diagonal entries of a matrix.
\newpage

For a matrix $\mathbf{Z} \in \mathbb{C}^{n \times d}$, the orthogonal projection onto this tangent space is given by
\begin{equation}\label{eq:oblique_proj}
\mathcal{P}_\mathbf{X}(\mathbf{Z}) = \mathbf{Z} - \mathbf{X} \cdot \operatorname{diag}(\mathbf{X}^\dagger \mathbf{Z}),
\end{equation}
ensuring that the result lies in $\mathcal{T}_\mathbf{X}(\mathcal{OB})$.

Equipping the tangent space with a Riemannian metric creates a \emph{Riemannian manifold}, which in turn unlocks differential geometry tools for optimization. 
A straightforward metric is the canonical Euclidean inner product, given by
\begin{equation}\label{eq:riem_metric}
\langle \mathbf{Z}_1, \mathbf{Z}_2 \rangle_\mathbf{X} = \operatorname{Re}\left( \operatorname{Tr}(\mathbf{Z}_1^\dagger \mathbf{Z}_2) \right).
\end{equation}

This metric allows for the definition of gradients, retractions, and other optimization elements directly on the manifold, enabling efficient, intrinsic constrained optimization without the need for explicit constraint handling.
In optimization over the sphere $S^{n-1}$, a point $\mathbf{X}$ is updated along a direction $\mathbf{\Xi}$ (e.g., in steepest descent) using a retraction to remain on the manifold, $i.e$
\begin{equation}\label{eq:retraction}
\text{Retract}_{\mathbf{X}}(\mathbf{\Xi}) = \frac{ \mathbf{X} + \alpha \mathbf{\Xi} }{ \| \mathbf{X} + \alpha \mathbf{\Xi} \| },
\end{equation}
which ensures that the new iterate remains on $S^{n-1}$.

\subsection{Oblique Manifold as a Quantum State}
Per the notation in \cite{Smart2024manybody}, a vectorized matrix can be expressed as a maximally entangled quantum state
\begin{equation}\label{eq:vec}
|\bm{\Psi}\rangle = 2^{-\frac{d}{2}} \sum_{k=0}^{d-1} |k\rangle \otimes |\psi_k\rangle = 2^{-\frac{d}{2}} \operatorname{vec}(\mathbf{X}),
\end{equation}
where $|\psi_k\rangle \in \mathcal{H}^n$ ($n$-dimensional Hilbert Space) represent the columns of $\mathbf{X}$.

Storing $\mathbf{X} \in \mathbb{R}^{n \times d}$ in the form of $|\bm{\Psi}\rangle$ enhances storage efficiency. 
While classically $\mathbf{X}$ demands $nd$ real numbers—which becomes unwieldy for large $n$ and $d$—the state $|\bm{\Psi}\rangle$ requires only $\lceil \log_2 d \rceil + \lceil \log_2 n \rceil$ qubits, thus offering exponential savings. 
Although quantum measurement limits direct access to individual elements, this compact representation supports efficient quantum algorithms for linear algebra and manifold optimization, with promising applications in advanced signal processing.

\subsection{Quantum Manifold Optimization}
Gradient descent, as a fundamental iterative minimization method, extends naturally to Riemannian manifolds for the purpose of optimizing over curved spaces. 
Remarkably, these algorithms can be implemented on quantum computers by mapping elements such as retractions, gradients, Hessians, and vector transports to quantum circuits via maximally entangled states. 
This approach leverages the logarithmic qubit encoding of high-dimensional manifolds, promising quantum-enhanced optimization capabilities.

For example, the oblique manifold can be adapted for use within a quantum circuit.
Given a function $f$, the Riemannian gradient is obtained by projecting the Euclidean gradient onto the tangent space
\begin{equation}\label{eq:grad}
\operatorname{grad} f(\mathbf{X}) = \mathcal{P}_\mathbf{X}(\operatorname{egrad} f(\mathbf{X})),
\end{equation}
which, in quantum-algebraic terms, corresponds to
\begin{equation}\label{eq:quant_proj}
\mathcal{P}_\mathbf{X}(|\bm{Z}\rangle) = \sum_{k=0}^{d-1} |k\rangle \left( |z_k\rangle - |\psi_k\rangle \langle \psi_k | z_k \rangle \right).
\end{equation}

In addition to projections, retractions that map tangent vectors to the manifold can be efficiently represented utilizing the quantum computing maximally entangled states with minimal qubit requirements.
Retractions facilitate navigation by approximating geodesic paths, such as those on $\mathcal{S}^{n-1}$, in a computationally tractable manner.
A retraction at a point $\mathbf{X} \in \mathcal{M}$ in the direction $\mathbf{V} \in \mathcal{T}_\mathbf{X} \mathcal{M}$ is defined by a function 
\begin{subequations}
\label{eq:retraction_def}
\begin{equation}
R_\mathbf{X}: \mathcal{T}_\mathbf{X} \mathcal{M} \to \mathcal{M},
\end{equation}
satisfying
\begin{equation}
R_\mathbf{X}(0) = \mathbf{X}, \quad \frac{d}{dt} R_\mathbf{X}(t \mathbf{V}) \Big|_{t=0} = \mathbf{V}.
\end{equation}
\end{subequations}

The retraction in (\ref{eq:retraction}) fulfills these conditions, although it is not unique.
For $\mathcal{S}^{n-1}$, another natural choice is the Riemannian exponential map \cite{boumal2023}
\begin{equation}\label{eq:exp_map}
R_{\mathbf{x}}(t \mathbf{v}_i) = \cos(t \|\mathbf{v}_i\|) \mathbf{x}_i + \sin(t \|\mathbf{v}_i\|) \frac{\mathbf{v}_i}{\|\mathbf{v}_i\|},
\end{equation}
which moves a point $\mathbf{x}_i \in \mathcal{S}^{n-1}$ (a column of matrix $\mathbf{X}$) along a geodesic in the direction $\mathbf{v}_i \in \mathcal{T}_{\mathbf{x}} \mathcal{S}^{n-1}$.

\begin{subequations}
\label{eq:matrix_form}
This motion is replicated using a matrix exponential suitable for quantum state representation.
A skew-symmetric matrix 
\begin{equation}
\mathbf{A}_i = \mathbf{v}_i \mathbf{x}_i^{\mathrm{T}} - \mathbf{x}_i \mathbf{v}_i^{\mathrm{T}}
\end{equation}
achieves this because $e^{t \mathbf{A}_i}$ is orthogonal, preserving norms and thereby ensuring that the result remains on the manifold.

In the orthonormal basis $\{ \mathbf{x}_i, \frac{\mathbf{v}_i}{\|\mathbf{v}_i\|} \}$, the matrix $\mathbf{A}_i$ simplifies to:
\begin{equation}
\mathbf{A}_i = \begin{pmatrix} 0 & -\|v_i\| \\ \|v_i\| & 0 \end{pmatrix}.
\end{equation}
which corresponds to a rotation matrix in the 2D subspace spanned by $\mathbf{x}_i$ and $\mathbf{v}_i$.
\end{subequations}

Consequently, the matrix exponential\footnote{Implementing matrix exponentials on quantum hardware remains an actively studied challenge, with ancilla-free approaches through Trotterization currently representing the most practical and widely adopted solution\cite{Childs_2021}.} is given by
\begin{equation}\label{eq:exp_matrix}
e^{t \mathbf{A}_i} = \begin{pmatrix} \cos(t \|v_i\|) & -\sin(t \|v_i\|) \\ \sin(t \|v_i\|) & \cos(t \|v_i\|) \end{pmatrix},
\end{equation}
which applyed to $\mathbf{x}_i$ yields
\begin{equation}\label{eq:exp_classical}
e^{t \mathbf{A}_i} \mathbf{x}_i = \cos(t \|v_i\|) \mathbf{x}_i + \sin(t \|v_i\|) \frac{\mathbf{v}_i}{\|v_i\|},
\end{equation}
therefore matching (\ref{eq:exp_map}).

For the oblique manifold, each column $\psi_k \in \mathcal{S}^{n-1}$ is updated via a column-wise exponential map
\begin{equation}\label{eq:quantum_retraction}
\text{R}_\mathbf{X}(|\mathbf{V}\rangle) = \sum_{k=0}^{d-1} |k\rangle \otimes e^{t \mathbf{A}_k} |\psi_k\rangle,
\end{equation}
where this quantum operator efficiently encodes the manifold’s geometry, thus minimizing the required qubit count.

\begin{algorithm}[H]
\caption{Summary of Quantum Manifold Optimization}
\begin{algorithmic}[1]
\STATE \textbf{Initialize Qubits}: Prepare all qubits in the $|0\rangle$
\STATE \textbf{Apply Quantum Gates}: 
\STATE \quad Apply Hadamard gates to create superpositions.
\STATE \quad Apply iteratively parameterized unitaries via retractions and projections,
\STATE \textbf{Execute Circuit}: Run the quantum circuit on the quantum processor.
\STATE \textbf{Measure Qubits}: Perform measurements in the computational basis to obtain classical outcomes.

\end{algorithmic}
\end{algorithm}

Finally, the inner product defined in (\ref{eq:riem_metric}) can be represented in terms of quantum states using the trace properties
\begin{equation}\label{eq:quant_inner}
\langle \mathbf{Z}_1, \mathbf{Z}_2 \rangle = \operatorname{Re}\big(\operatorname{vec}(\mathbf{Z}_1)^\dagger \operatorname{vec}(\mathbf{Z}_2) \big),
\end{equation}
which can be similarly expressed as:
\begin{equation}\label{eq:quant_trace}
\langle \mathbf{Z}_1, \mathbf{Z}_2 \rangle = 2^{d} \langle \mathbf{\Psi}_{Z_1} | \mathbf{\Psi}_{Z_2} \rangle.
\end{equation}
As long as objectives (typically representing total energy, power, or other quantities) can be expressed in this form, the methods outlined above enable solving the manifold problem using quantum circuits.

\vspace{-1ex}
\section{Applications of \ac{QMO} in Wireless Systems}

In this section, several relevant applications in wireless system design that are classically defined on manifolds are considered, and the corresponding manifold representations reformulated under the \ac{QMO} framework are derived.
Specifically, we elaborate in detail with the first couple of examples: pilot sequence design for CF-mMIMO scenarios, which forms an oblique manifold optimization problem, and beamformer design for gigantic MIMO, which constitutes a Stiefel manifold optimization problem.
In addition, we consider several other applications and their reformulations, such as the \ac{RIS} phase-optimization and various \ac{ISAC}-related problems \cite{ranasinghe2025}.

\vspace{-1ex}
\subsection{Pilot Sequence Design for CF-\acs{mMIMO}}
Distributed MIMO systems have attracted significant attention, with \ac{CF}-\acs{mMIMO} being a prominent example where many access points (APs) spread over a large area serve multiple users simultaneously via fronthaul-connected \ac{CPU} \cite{Ngo_2017}.
One critical challenge in such systems is pilot contamination---the situation when the number of users exceeds the number of available orthogonal pilot sequences, degrading channel estimation quality.
To counter pilot contamination, pilot sequences can be designed using a manifold optimization framework, aiming to derive nearly orthogonal sequences that minimize the contamination/mutual interference.

In a system with $L$ APs, $K$ users, and $T$ orthogonal sequences (with $K > T$), the design is cast as
\begin{equation}
    \begin{aligned}
        \underset{ \bar{\bm{f}}_k \in\mathbb{C}^{T\times 1} }{\text{minimize}} \quad &  \sum_{\ell=1}^L \sum_{k=1}^K \sum_{k'\in\mathcal{K}\backslash\{k\}} \beta_{\ell,k'}\left| \bar{\bm{f}}_k^\dagger \bar{\bm{f}}_{k'}\right|^2 \\
        \mathrm{s.t.} \quad & \|  \bar{\bm{f}}_k\|_2^2 = 1, \quad \forall k,
    \end{aligned}
\end{equation}
where $\beta_{\ell,k'}$ denote the large-scale fading coefficients, and $\bar{\bm{F}} \in \mathbb{C}^{T \times K}$ is the pilot matrix.

After some trivial algebra, the problem can be reformulated as an oblique manifold problem, namely
\begin{equation}
    \min_{\bar{\bm{f}}_k \in \mathcal{OB}(T,K)} \ \mathrm{Tr}\Big(\bm{B} \, \lvert \bar{\bm{F}}^\dagger \bar{\bm{F}} \rvert^2 \mathbf{A}\Big),
\end{equation}
where $\bm{B}$ collects the fading coefficients and $\mathbf{A} \in \mathbb{R}^{K \times L}$ is a matrix of ones.

As discussed in the previous section, a quantum perspective can be introduced by vectorizing $\bar{\bm{F}}$ into a normalized quantum state
\begin{equation}
    \vert \bm{\Psi} \rangle = 2^{-K/2} \sum_{k=0}^{K-1} \vert k \rangle \otimes \vert \psi_k \rangle,
\end{equation}
which embeds the pilot matrix into the quantum domain.

The trace objective can then be recast as
\begin{equation}
\begin{aligned}
    &\operatorname{Tr}\Big((\bar{\bm{F}}^\dagger \bm{B} \bar{\bm{F}})(\bar{\bm{F}}^\dagger \mathbf{A} \bar{\bm{F}})\Big) =\\ &2^{2K} \sum_{i,k} \langle \bm{\Psi} \vert \mathbf{M}_{ki} \otimes \bm{B} \vert \bm{\Psi} \rangle \langle \bm{\Psi} \vert \mathbf{M}_{ik} \otimes \mathbf{A} \vert \bm{\Psi} \rangle.
    \end{aligned}
\end{equation}
where $M_{ki} = |k\rangle\langle i|$.

This quantum encoding facilitates the use of quantum circuits for efficient conjugate gradient descent, where the gradient can similarly be represented\footnote{QMO has been shown to perform effectively in simulations for the Many-Body Eigenstate Problem using Qiskit. Nonetheless, further validation through implementation on an actual quantum computer is required to fully assess its practical efficacy.}.

\subsection{Beamformer Design for Gigantic \ac{MIMO}}
Gigantic \ac{MIMO} extends conventional massive \acs{MIMO} by leveraging an extremely large number of antennas at the transmitter and/or receiver, thereby enhancing spectral and energy efficiencies through improved spatial resolution and diversity \cite{Bjornson_2024}.

The basic system model for the received signal is given by
\begin{equation}
    \mathbf{y} = \mathbf{H} \mathbf{x} + \mathbf{n},
\end{equation}
with $\mathbf{H} \in \mathbb{C}^{n_r \times n_t}$ as the channel matrix, $\mathbf{x} \in \mathbb{C}^{n_t}$ the transmitted signal, and $\mathbf{n} \sim \mathcal{CN}(0, \sigma^2 \mathbf{I}_{n_r})$ representing noise.

When applying beamforming, a precoding matrix $\mathbf{W} \in \mathbb{C}^{n_t \times n_r}$ is used, modifying the signal model to
\begin{equation}
    \mathbf{y} = \mathbf{H} \mathbf{W} \mathbf{x} + \mathbf{n}.
\end{equation}

Here, $\mathbf{W}$ is chosen such that its columns are orthonormal, directing beams towards intended users while suppressing interference.
The beamforming design problem is formulated to maximize the received power
\begin{equation}
    \begin{aligned}
        \underset{\mathbf{W} \in \mathbb{C}^{n_t \times n_r}}{\text{max}} \quad & f(\mathbf{W}) = \operatorname{Tr}\Big(\mathbf{W}^\dagger \mathbf{H}^\dagger \mathbf{H} \mathbf{W}\Big) \\
        \text{s.t.} \quad & \mathbf{W}^\dagger \mathbf{W} = \mathbf{I}_{n_r},
    \end{aligned}
\end{equation}
ensuring that $\mathbf{W}$ lies on the Stiefel manifold (the set of matrices with orthonormal columns).

In the quantum framework, the beamforming matrix is encoded by vectorization, $i.e$
\begin{equation}
    \operatorname{vec}(\mathbf{W}) = \sum_{k=0}^{n_t-1} \vert k \rangle \otimes \vert \psi_k \rangle,
\end{equation}
which leads to a reformulation of the cost function as the expectation value of a Hermitian operator
\begin{equation}
    f(\mathbf{W}) = 2^{n_t} \langle \Psi \vert \mathbf{I}_{n_r} \otimes \mathbf{R} \vert \Psi \rangle,
\end{equation}
where $\mathbf{R} = \mathbf{H}^\dagger \mathbf{H}$.

Riemannian optimization methods, such as Riemannian gradient descent, can then be employed to efficiently traverse the Stiefel manifold, thereby harnessing the structured geometry inherent in the quantum encoding of the beamforming problem.

\subsection{Optimization of Reconfigurable Metasurface Parameters}

Consider an \ac{RIS} composed of $N$ unit-modulus reflecting elements. Let $\boldsymbol{\theta} \in \mathbb{R}^N$ denote the phase shifts applied at each element. 
The effective channel from the transmitter to the receiver via the \ac{RIS} is modeled as
\begin{equation}
\mathbf{h}_{\text{eff}} = \mathbf{h}_r^\dagger \mathbf{\Phi}(\boldsymbol{\theta}) \mathbf{H}_b \mathbf{f},
\end{equation}
where $\mathbf{h}_r \in \mathbb{C}^{N}$, $\mathbf{H}_b \in \mathbb{C}^{N \times M}$, $\mathbf{f} \in \mathbb{C}^{M}$, and $\mathbf{\Phi}(\boldsymbol{\theta}) = \mathrm{diag}(e^{j\theta_1}, \ldots, e^{j\theta_N})$.

The corresponding optimization problem is then
\begin{equation}
\max_{\boldsymbol{\theta} \in [0, 2\pi)^N} |\mathbf{h}_{\text{eff}}|^2 = \max_{\boldsymbol{\theta}} \left| \mathbf{h}_r^\dagger \mathbf{\Phi}(\boldsymbol{\theta}) \mathbf{H}_b \mathbf{f} \right|^2,
\end{equation}
which is a manifold optimization over the torus $ (S^1)^N $, $i.e.$, a product of unit circles, which following similar approaches to the previous examples, can be solved under the \ac{QMO} technique, by expressing the objective as
\begin{equation}\label{eq:RIS_trace}
|\mathbf{h}_{\text{eff}}|^2 = \operatorname{Tr}\Bigl(
\mathbf{\Phi}(\boldsymbol{\theta})\, \mathbf{Q}\, \mathbf{\Phi}^\dagger(\boldsymbol{\theta})\, \mathbf{R}
\Bigr).
\end{equation}
where $\mathbf{Q} \triangleq \mathbf{H}_b \mathbf{f}\,\mathbf{f}^\dagger \mathbf{H}_b^\dagger$ and $\mathbf{R} = \mathbf{h}_r \mathbf{h}_r^\dagger$.

The objective (\ref{eq:RIS_trace}) can be mapped to a maximally entangled quantum state, similar to previous examples, by representing the matrix through vectorization.

\vspace{-0.85ex}
\subsection{Other Relevant Problems}

Trivially, there is a plethora of other relevant optimization problems in wireless communications and signal processing over manifolds which can be framed under similar procedure provided in the examples above, which are not elaborated due to space limitations -- but are equally critical and effective.
To list a few, such applications include waveform design for optimal sensing behaviors which has been shown require unit-modulus signal coefficients \cite{Rexhepi_arxiv2024}, other waveform constraints over Doppler-shift-dependent doubly-dispersive channels \cite{Rou_SPM24} especially for detection of next-generation signals over unitary and chirp-like domains \cite{Rou_Asilomar24}, or the estimation of environment scatters in \ac{ISAC} regimes which can be captured via a reflective phase-shift of the impinging signal \cite{Rou_TWC24}.

\section{Conclusion}

In this paper, we have introduced \ac{QMO} as a foundational framework for the design of wireless communication systems under geometric constraints, bearing in mind that the scale of \ac{B6G} systems is likely to grow exponential due to trends such as the proliferation of \ac{RIS} \cite{Chepuri2023integrated}, the implementation of large-scale \ac{CF}-\ac{mMIMO} architectures \cite{Ngo_2017}, and the deployment of gigantic MIMO \cite{Bjornson_2024} structures. 
By leveraging structured manifolds such as the Stiefel, Grassmannian, and oblique manifolds, we demonstrate how several canonical wireless design problems can be reformulated in terms of trace-based objectives suitable for quantum manifold optimization. 
Variational encodings and quantum geometric tools are derived and presented to perform Riemannian optimization directly on quantum devices.
Our examples -- including pilot design for \ac{CF}-\ac{mMIMO}, beamformer optimization for gigantic \ac{MIMO}, and \ac{RIS} phase configuration -- illustrate the wide applicability of the approach in wireless systems. 
We prospect and stimulate future works to focus on simulation studies, circuit designs, and performance benchmarks against classical solvers.

\end{document}